\documentclass[12pt]{PoS}
\newcommand{\pT}{p_T}
\title{W asymmetries at CDF and D0 }
\ShortTitle{W asymmetries at CDF and D0 }

\author{\speaker{Heidi Schellman}\thanks{For the D0 and CDF collaborations}\\
        Northwestern University\\
        E-mail: \email{h-schellman@northwestern.edu}}

\abstract{We present recent  $W$ and charged lepton asymmetry measurements from the CDF and D0 experiments.
Theoretical predictions agree with the  CDF $W$ asymmetry, measured using a new matrix element technique. These theoretical predictions are less consistent with the latest lepton asymmetry measurements from D0 and CDF,
especially for high charged lepton transverse momentum. 
}

\FullConference{Submitted to the XVIII International Workshop on Deep-Inelastic Scattering and Related Subjects, DIS 2010,
		April 19-23, 2010
		Firenze, Italy\\
		{\it This preprint has additional figures not included in the conference proceedings.}}

\begin{document}

\section{Introduction}
\newcommand{\antip}[0]{{\overline{p}}}
\newcommand{\ppbar}{p\antip}
\newcommand{\infb}{fb$^{-1}$}
\newcommand{\quark}{\qu{q}}
\newcommand{\antiquark}{\overline{\quark}}
\newcommand{\uquark}{\qu{u}}
\newcommand{\dquark}{\qu{d}}
\newcommand{\squark}{\qu{s}}
\newcommand{\cquark}{\qu{c}}
\newcommand{\bquark}{\qu{b}}
\newcommand{\tquark}{\qu{t}}
\newcommand{\antiuquark}{\aqu{u}}
\newcommand{\antidquark}{\aqu{d}}
\newcommand{\antisquark}{\aqu{s}}
\newcommand{\anticquark}{\aqu{c}}
\newcommand{\antibquark}{\aqu{b}}
\newcommand{\antitquark}{\aqu{t}}
\newcommand{\tbar}{\overline{t}}
\newcommand{\bbar}{\overline{b}}
\newcommand{\cbar}{\overline{c}}
\newcommand{\dbar}{\overline{d}}
\newcommand{\ubar}{\overline{u}}
\newcommand{\sbar}{\overline{s}}
\newcommand{\gt}{\rightarrow}
\newcommand{\etal}{{\it et al.}}

$W$ bosons are produced in $\ppbar$ collisions via the reactions $u\dbar \gt W^+$ and $d\ubar\gt W^-$.
  If one makes the reasonable assumption that the anti-quark content of the anti-proton is the same as the quark content of the proton, any difference between the $u$ quark and $d$ parton momentum distributions in the proton will result in an asymmetry
  in $W$ boson production as a function of the longitudinal momentum of the $W$, $p_z(W) = (x_p - x_\antip) \sqrt{s}/2$,
   where $x_p$ and $x_\antip$ are the momentum fractions carried by the partons in the proton and anti-proton and $\sqrt{s}$ is the $p\antip$ center-of-mass energy.  In a very simple model, where quarks in the proton dominate, the $W$ boson production
  asymmetry 
  
  \begin{eqnarray*}\label{eq:definition}
  A(W) &=&{ {N(W^+) - N(W^-)}\over {N(W^+) + N(W^-)}} \\&\simeq& {{u(x_p) d(x_\antip) - d(x_p) u(x_\antip)}\over{
   u(x_p) d(x_\antip) + d(x_p) u(x_\antip)}}
   \end{eqnarray*}
   
\begin{figure}
\begin{center}
\includegraphics[width=.8\textwidth]{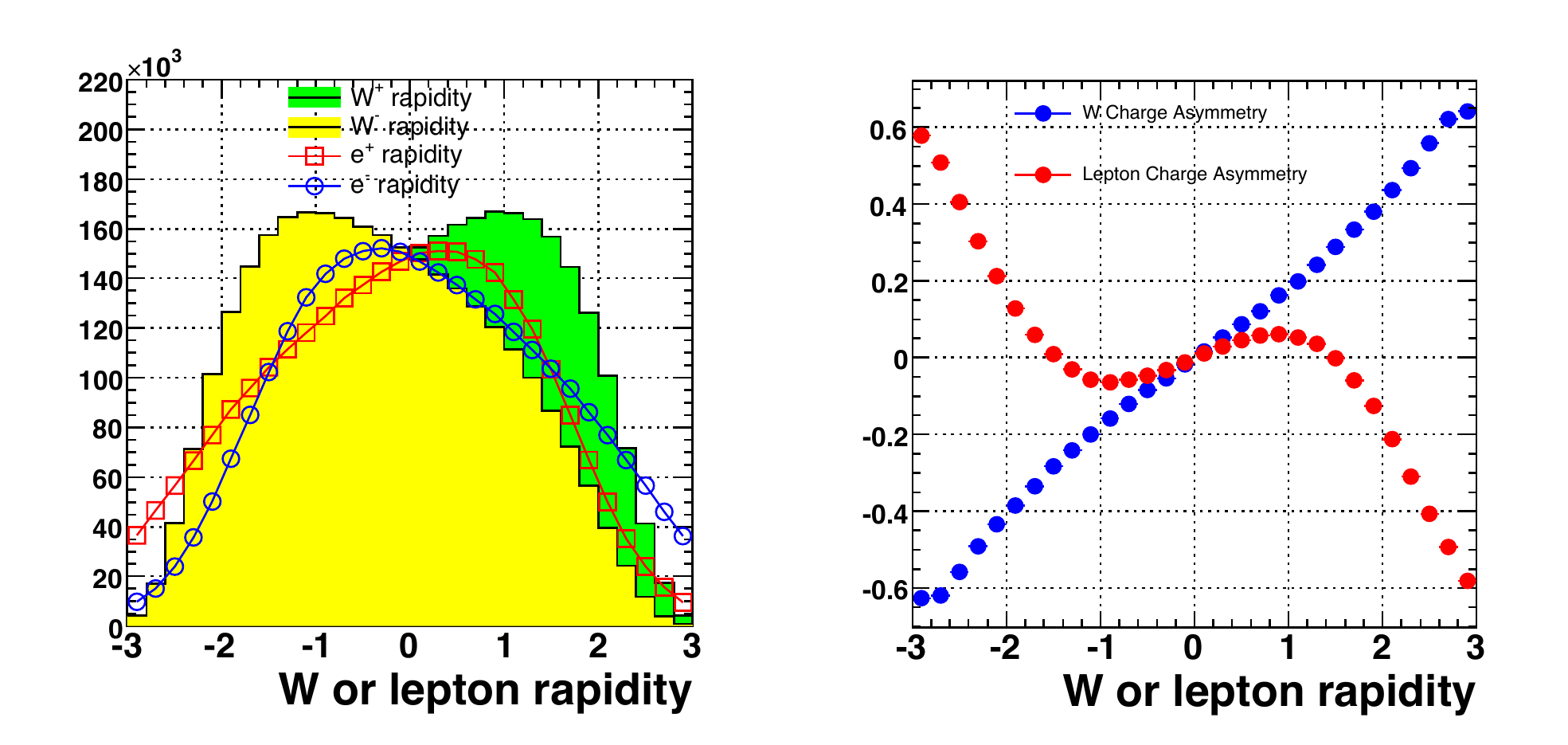}
\end{center}
\caption{ (left) Rapidity distribution of $W$ bosons and decay leptons produced at $\sqrt{s} = 1960$ GeV. (right) The resulting asymmetry for $W$ bosons and $leptons$. }
\label{fig1}
\end{figure}

    This simple partonic interpretation is for illustrative purposes  only as it is valid in the limit where anti-quarks in the proton do not contribute and where higher order QCD contributions are negligible.  The $W$  asymmetry  can still be used to constrain parton distribution functions in the presence of QCD but a full production and decay model must be used at the appropriate order in perturbation theory.
  $W$ bosons are most easily detected via their leptonic decays $W\gt \ell\nu_\ell$, in particular the $e\nu_e$ and $\mu\nu_\mu$ final states. Due to the $V-A$ nature of the decay, the charged lepton tends to be oriented opposite to the $W$ direction, which washes out the initial $W$ asymmetry.  Figure~\ref{fig1} illustrates the production rates and asymmetries for leptons and $W$'s as a function of rapidity, and pseudorapidity, $$y(W)    =     \frac{1}{2} \ln \left(\frac{E+p_z}{E-p_z}\right), \qquad\qquad \eta(\ell) =      \frac{1}{2} \ln \left(\frac{p+p_z}{p-p_z}\right) =      -\ln\tan\left(\frac{\theta}{2}\right).$$

  We present two measurements of asymmetries from the CDF and D0 experiments at the Tevatron.  The CDF experiment has concentrated on measurement of the $W$ asymmetry, using a new and sophisticated matrix element reweighting technique\cite{CDFmethod,CDFresult} with a $W$ mass constraint to constrain the decay neutrino phase space.  The D0 collaboration uses a more traditional method\cite{D0e,D0mu,CDFe}, where the leptonic asymmetry is measured and compared with theoretical predictions to extract the parton distributions.  The CDF $W$ measurement requires some theoretical assumptions about $W$ production and decay in the extraction of the asymmetry while the leptonic asymmetry is almost purely an experimental number with few theoretical assumptions.   The CDF collaboration have recently done an extraction of the leptonic asymmetries using the same data sample as the W asymmetry but using cuts similar to those used by D0. This allows a direct
comparison of the experimental data between the two experiments.

\section{Measurement of the W asymmetry at CDF}

\begin{figure}
\begin{center}
\includegraphics[width=.6\textwidth]{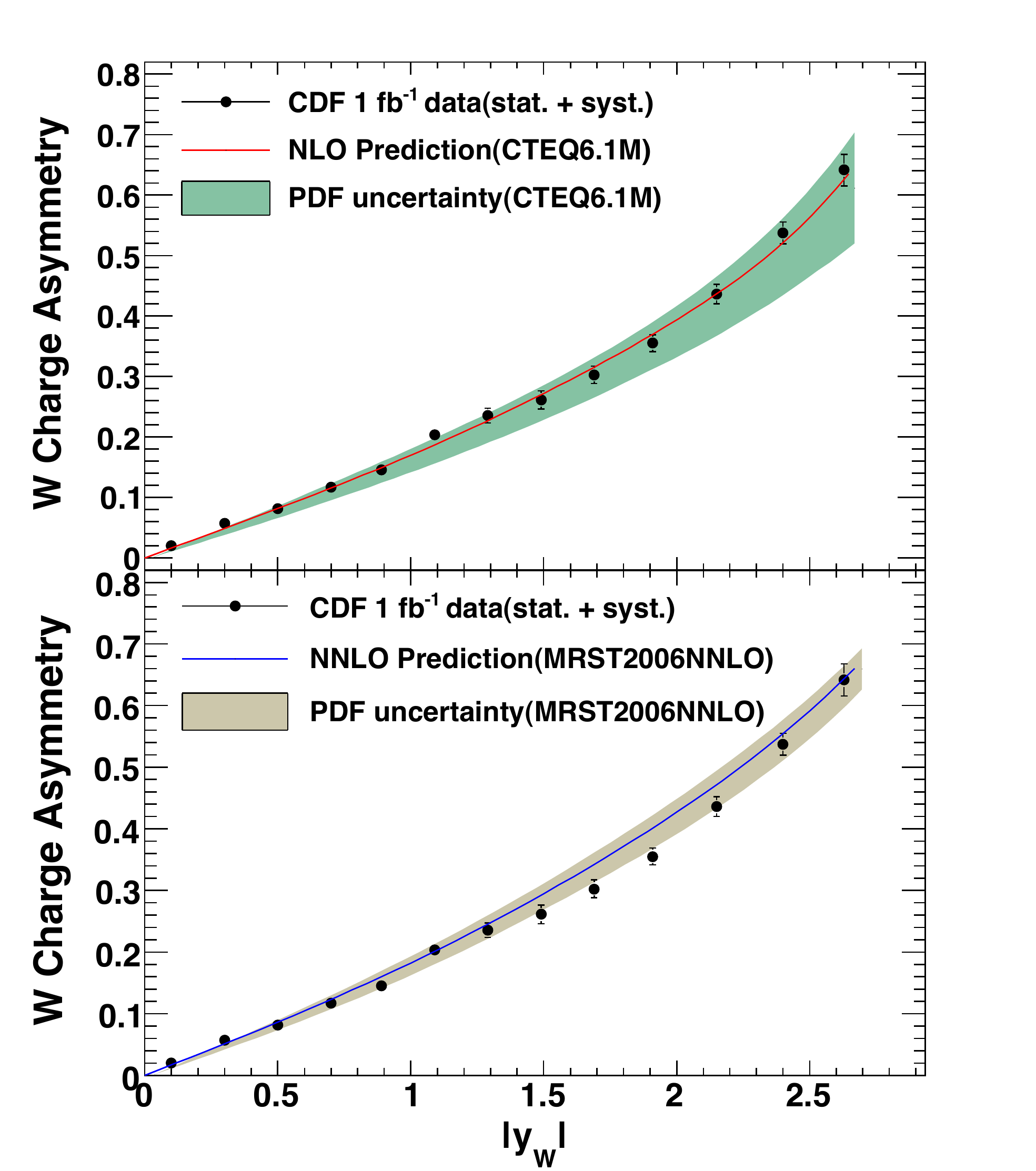}
\end{center}
\caption{ (top)  CDF measurement of the $W$ boson asymmetry, compared to the CTEQ6.6M parton distribution set.
(bottom) CDF measurement of the $W$ boson asymmetry, compared to the MRST2006 parton distributions. }
\label{fig 2}
\end{figure}

The CDF collaboration have made a direct measurement of the asymmetry of the $W$ boson by finding the two neutrino longitudinal momentum solutions which, when combined with an electron, reconstruct to the $W$ mass.   Both solutions are considered but a matrix element method is used to calculate the  likelihood of the two solutions and that likelihood is then used as an event weight in the asymmetry estimate.  The likelihood is estimated
from simulation and includes the \bf unsigned \rm $W$ boson rapidity distribution and the  relative contribution of quarks and antiquarks to the $W$ boson decay angular distribution but explicitly does not include asymmetry information.  To avoid any 
residual influence of the asymmetry assumptions in the simulation, the theoretical model is iterated.  Figure~\ref{fig 2} shows the CDF $W$ asymmetry result compared to both the CTEQ6.1M\cite{CTEQ61,MCNLO} and the MRST2006\cite{MRST2006, NNLO} parton distribution sets.

\section{Measurement of the muon asymmetry in $W$ production at D0}

\begin{figure}
\begin{center}
\includegraphics[width=.6\textwidth]{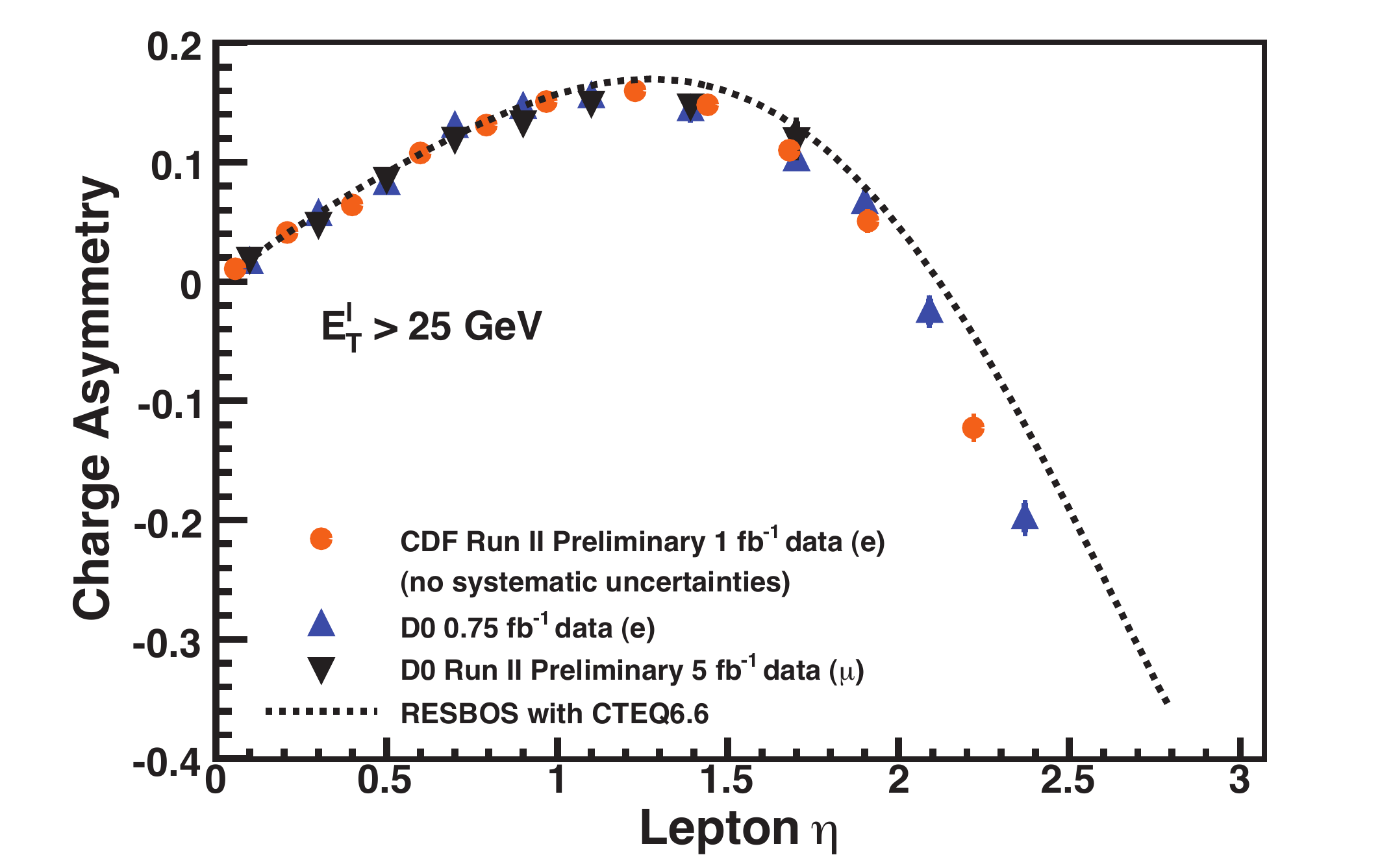}
\includegraphics[width=.6\textwidth]{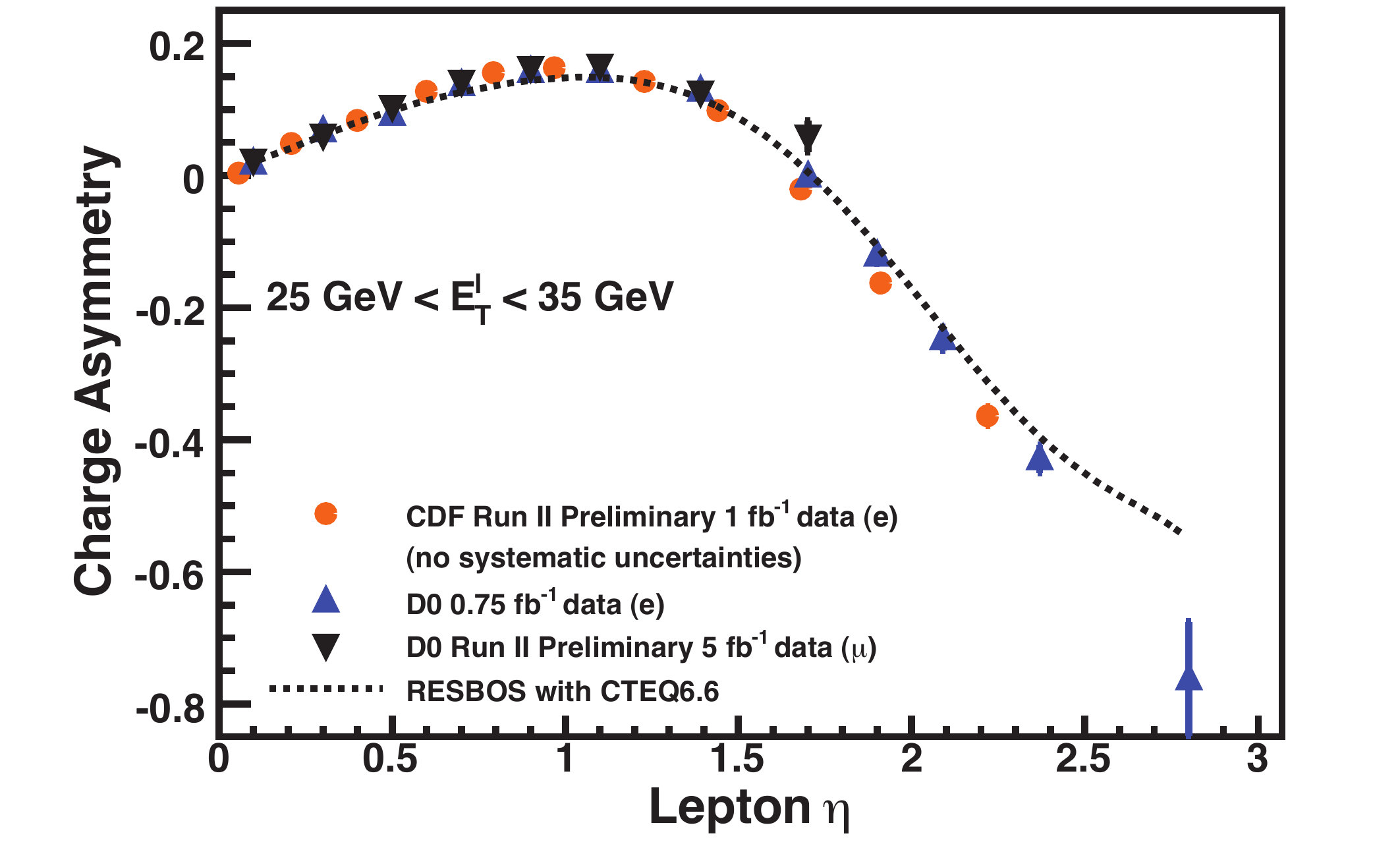}
\includegraphics[width=.6\textwidth]{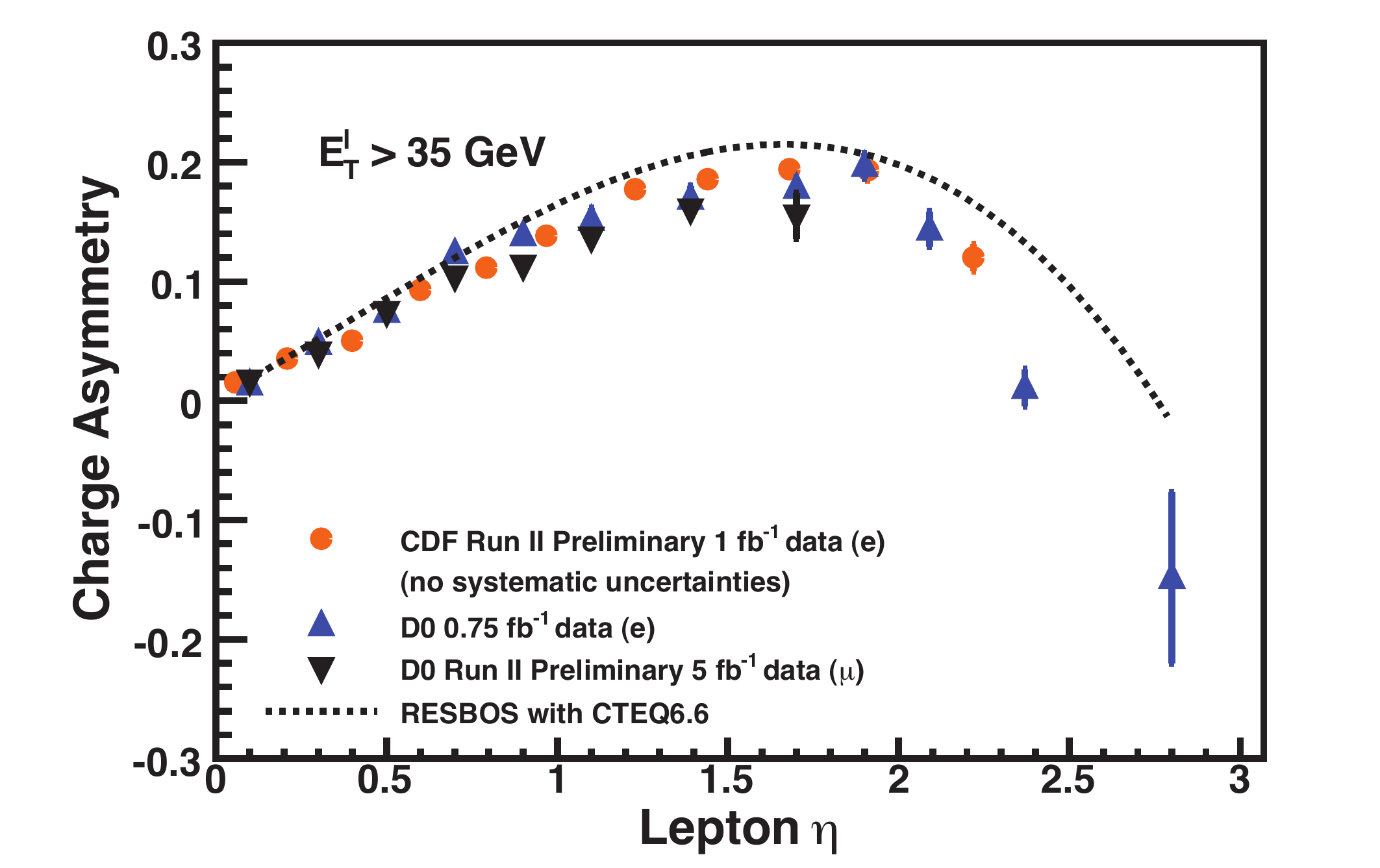}
\end{center}
\caption{(top) The D0 $\mu$ asymmetry compared to an earlier D0 measurement in the electron channel and to a preliminary CDF extraction of the lepton asymmetry from the same data as the $W$ asymmetry.  The curve shows the CTEQ6.6M parton distribution predictions. The lepton  and neutrino $\pT$ are both
required to be $> 25 $ GeV to allow comparison with the electron data. (middle)  A subset of these data with the $\mu$ $\pT$ required to be in the range $25 < \pT(\ell) < 35$ GeV. (bottom)  The subset of the data with  $\pT(\ell)$  $> 35$ GeV.}
\label{fig 3}
\end{figure}

The D0 experiment has recently made a preliminary measurement of the muon asymmetry in $W$ production using 4.9 \infb of integrated luminosity.  This data sample includes 2.3 million reconstructed $W$ boson decays.  $W$ boson candidates are required to have a reconstructed muon with transverse momentum $\pT(\mu)$ > 20 GeV and pseudorapidity $\eta(\mu) < 2.0$ and missing transverse energy MET$\simeq\pT({\nu}) > 20$ GeV.  Backgrounds to the $W$ signal are QCD jets misidentified as muons ,2.4\%, and $Z$ boson decays with one of the muons lost, 3.1\%.  The similar process $W \gt \tau\nu_\tau$ where the $\tau$ decays to $\mu+X$ contributes 3\%.
Charge misidentification is studied in $Z$ boson decays and is found to be approximately one in ten-thousand muons.  The raw lepton asymmetry is measured as a function of the lepton pseudorapidity.  The data presented have been corrected for backgrounds, the small charge misidentification rate and for momentum smearing using a full simulation of the D0 detector.  Figure~\ref{fig 3} shows the
data compared to theoretical calculations made using the NLO predictions with resummation from RESBOS\cite{RESBOS} and the 
CTEQ6.6M\cite{CTEQ66} parton distribution functions.  The data are shown for several charged lepton momentum bins and compared to previously published D0 data in the electron channel and a preliminary electron asymmetry extracted by CDF\cite{CDFnew} from the
same sample as the $W$ asymmetry discussed earlier.

For the inclusive sample, with $\pT(\ell,\nu) > 25$ GeV, the data with symmetric cuts on the neutrino and charged lepton $\pT$'s show good agreement with each other but lie above the theoretical prediction.  Remember that the $W$ asymmetry extracted from the same CDF sample was fully consistent with these predictions. 
If we subdivide the data into low ($25 < \pT(\ell) < 35$ GeV) and high ($\pT(\ell) > 35$ GeV) samples, we find that the data remain reasonably consistent but the theoretical predictions underestimate the asymmetry for for $|\eta| < 1.5$ in the low $\pT(\ell)$ sample and overestimate the asymmetry for the high $\pT(\ell)$ sample.

\section{Discussion}
\begin{figure}
\begin{center}
\includegraphics[width=.6\textwidth]{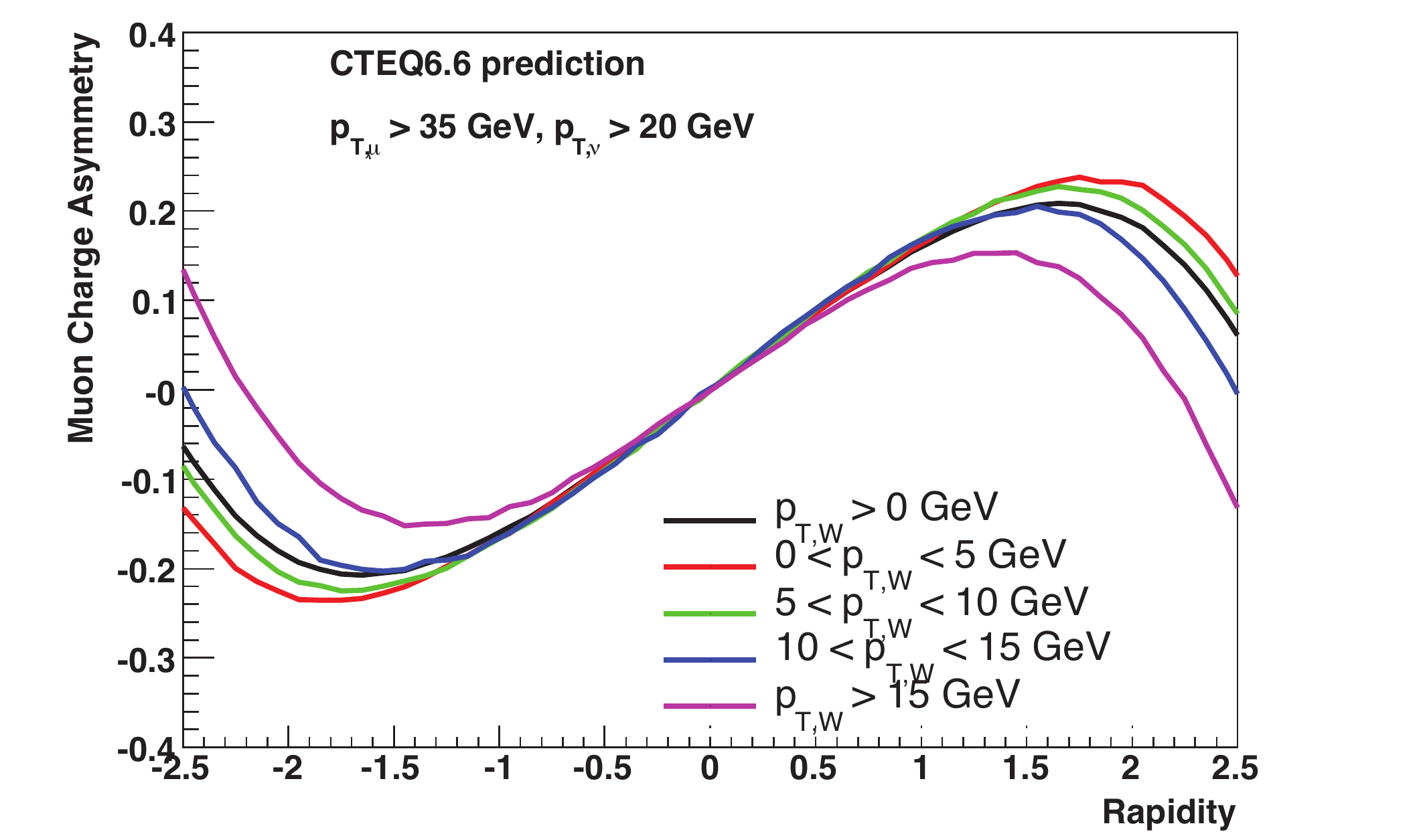}
\end{center}
\caption{RESBOS + CTEQ6.6M predictions for the lepton asymmetry with $\pT(\nu) > 20 $ GeV and $\pT(\mu) > 35$ GeV for several
values of the $W$ boson $\pT$.}
\label{fig 4}
\end{figure}

We are left with a conundrum, the $W$ asymmetry measurement from CDF agrees well with theoretical predictions, while the lepton asymmetries from CDF and D0 are reasonably consistent with each other but differ from theoretical expectations. One hint in the resolution of this problem may be the observation that the discrepancy is larger for the subset of events where $\pT(\nu) > 25$  and
$\pT(\ell) > 35$ GeV.  These asymmetric selections bias the sample towards events where the $W$ $\pT$ is non-zero.  A theoretical study by the D0 experiment\cite{D0newmu} of the $\pT(W)$ dependence of the lepton asymmetry is shown in Figure~\ref{fig 4} and confirms that the lepton asymmetry is quite sensitive to the $\pT (W)$ distribution, either through the $\pT(W)$ itself, or through helicity effects which may not be fully reproduced in the present simulations.  The $W$ asymmetry measurement, although it is also has some sensitivity
to theoretical models,  should be much less sensitive to the $\pT(W)$ and helicity modeling.

To conclude, the CDF and D0 collaborations have measured the $W$ and leptonic asymmetries in $\ppbar$ scattering using two
methods and two leptonic channels.  The data for leptonic asymmetries are reasonably consistent between channels and experiments but current theoretical predictions do not reproduce the leptonic asymmetries even though they do agree with the $W$ boson asymmetry.  

\end{document}